*Daniel Klug, Elke Schlote (2021): Designing a Web Application for Simple and Collaborative Video Annotation That Meets Teaching Routines and Educational Requirements. In: Proceedings of the 19th European Conference on Computer-Supported Cooperative Work: The International Venue on Practice-centred Computing on the Design of Cooperation Technologies, Reports of the European Society for Socially Embedded Technologies (ISSN 2510-2591), DOI: 10.18420/ecscw2021_ep27*

# Designing a Web Application for Simple and Collaborative Video Annotation That Meets Teaching Routines and Educational Requirements

Daniel Klug[1], Elke Schlote[2]
[1] Carnegie Mellon University, *dklug@cs.cmu.edu*
[2] University of Basel, *elke.schlote@unibas.ch*

**Abstract.** Video annotation and analysis is an important activity for teaching with and about audiovisual media artifacts because it helps students to learn how to identify textual and formal connections in media products. But school teachers lack adequate tools for video annotation and analysis in media education that are easy-to-use, integrate into established teaching organization, and support quick collaborative work. To address these challenges, we followed a design-based research approach and conducted qualitative interviews with teachers to develop TRAVIS GO, a web application for simple and collaborative video annotation. TRAVIS GO allows for quick and easy use within established teaching settings. The web application provides basic analytical features in an adaptable work space. Key didactic features include tagging and commenting on posts, sharing and exporting projects, and working in live collaboration. Teachers can create assignments according to grade level, learning subject, and class size. Our work contributes further insights for the CSCW community about how to implement user demands into developing educational tools.

# 1 Introduction

Audiovisual media artifacts, such as music videos or YouTube clips, are an essential part of teenagers' and younger adults' everyday life and their primary means to develop digital media literacy (Lange, 2016). In line with the dramatic increase in media consumption in recent decades by kids and teenagers (McNally and Harrington, 2017), school curricula more and more require didactic approaches to teach the analysis of audiovisual media artifacts (BMBWF, 2019; Erziehungsdepartement Basel-Stadt, 2013; Ministerium für Kultus, Jugend und Sport Baden-Württemberg, 2016).

In this context, teachers are ultimately the decision makers when it comes to choosing educational tools and how to integrate them into existing teaching practices. Working with digital educational tools means teachers need to provide digital media content, design innovative learning environments, and develop a collaborative environment to engage with students in an effort to improve peer learning and teacher-student-interaction. While audiovisual media artifacts are popular learning resources in *Language* classes, *Music*, or *Arts*, teachers lack adequate digital educational tools to teach video analysis and annotation according to these new curricula standards. Teachers commonly use apps like GarageBand or iMovie, which are not primarily designed for educational purposes. Meanwhile, studies find that teachers are reluctant to use advanced analytical tools (Chien et al., 2014; Sang et al., 2011) and do not always see educational benefit in using apps recommended by educational institutions (Al-Zaidiyeen et al., 2010). In addition, Holstein et al. (2017) show teachers stop using digital tools if they cannot adapt to changes in curricula, hamper tracking student performance and following their learning process, and allow students to trick the system rather than tackling the tasks.

The main observation is that many available tools are too complex or too limited. They are either not primarily designed for use in *Language*, *Arts*, or *Music* classes, or exclusively designed for specific subjects or learning activities. However, a more crucial problem is, that teachers are generally reluctant to use digital tools because analyzing audiovisual media artifacts is a rather minor aspect in most subjects and there is only little guideline for how to adapt digital tools into established teaching routines. To ensure better accessibility and usability for educators, educational tools for video annotation need to provide the right set of features but foremost tools need to mind the didactic and methodological context in which they should be used in the first place. For example, teaching time is scarce and not intended for lengthy introductions and explanations of tools; film and video analysis in secondary education draws on a well-defined and unchangeable set of analytical categories (Bordwell, 2004) but requires fewer advanced features than in the academe; teachers and students alike need a clear motivation and benefit to work with a tool work should be quick and easy to set up with a tool. Video annotation tools need to adapt to the didactic context rather than creating new ones. These tools should not require additional effort from teachers but present easy-to-use digital options for subject-specific activities. This results in a strong need for adequate educational video annotation tools.

This paper presents the development and design of TRAVIS GO, a web application for video analysis and annotation (http://travis-go.org/en) that meets (a) the educational and didactic requirements of curricula, and (b) the needs and demands of teachers for easy integration into teaching contexts (Schlote and Klug, 2020). Our work is guided by the following research questions:

- **RQ1:** *What are didactic requirements for teaching about audiovisual media artifacts?*
- **RQ2:** *What are features teachers need in digital educational tools to teach video annotation and analysis?*
- **RQ3:** *How can a digital educational tool for video annotation and analysis easily support collaboration and teacher-student interaction?*

To answer these research questions, we followed a design-based research (DBR) approach (Design-Based Research Collective, 2003). *First*, we researched educational needs and demands for video annotation tools by analyzing curricula, *second*, we conducted expert interviews with teachers. *Third*, we developed TRAVIS GO following an iterative process of designing and redesigning the web application.

# 2 Collaboration and Audiovisual Media Analysis and Annotation

Shorter audiovisual media formats, such as TV series, or music videos, are largely popularized by YouTube, TikTok, or Instagram. They present "video-mediated lifestyles" (Lange, 2016) for kids and teenagers whose digital media consumption (Frees et al., 2019; McNally and Harrington, 2017) and use of digital devices continues to steadily increase (Anderson and Jiang, 2018; Rideout and Robb, 2019). Studies demonstrate that students of all ages are highly familiar and socialized with various audiovisual media content (Medienpädagogischer Forschungsverbund Südwest, 2018; Suter et al., 2018). This further illustrates the need for teachers to be able to design didactic material that includes audiovisual media artifacts related to students' media life-worlds. To achieve this, they need adequate digital tools to realize cooperative learning in groups, to support individual work strategies, and to provide direct feedback (Bundesamt für Sozialversicherung, 2019).

Asynchronous collaboration in online contexts (Cadiz et al., 2000; Dorn et al., 2015; Weng and Gennari, 2004) as well as annotation practices to collaboratively work on (Bargeron et al., 2002; Diamant et al., 2008) and with media artifacts (Crabtree et al., 2004; Hartmann et al., 2010) are long-established research areas within CSCW. More recent work discusses and evaluates collaborative tools to improve distributed teamwork and group discussions, for example, through creating visual cues (Shi et al., 2017), visualizing group dynamics (Lim and Chiu, 2015), or sharing multimedia content in and after distributed team meetings (Marlow et al., 2016).

In educational contexts like computer-supported collaborative learning (CSCL), collaborative tools are designed, for example, to enable multimodal feedback (Yoon et al., 2016), improve students' skills to work in online groups (Ahuja et al., 2019), or to analyze the effects of collaborating in Google Docs on students' synchronous writing practices (Yim et al., 2017).

Previous work on collaboration and human-computer interaction presents various tools for annotating and analyzing media content (Burr, 2006; Cunha et al., 2013; Hosack, 2010; Liu et al., 2019) mostly designed for general (non-student) audiences (Hamilton et al., 2018; Nguyen et al., 2013). For example, tools to visualize Twitter claims (Pollalis et al., 2018), to analyze online information quality (Diakopoulos et al., 2009), to create dynamic annotations in web page text content (Hong and Chi, 2009), to automate video annotation (Wang et al., 2018), or to create multimodal annotations to improve creative processes in dance rehearsals (El Raheb et al., 2018; Singh et al., 2011).

From a learning science perspective, papers address how students learn with videos but not how they learn about videos. Papers, for example, analyze types of engagement with video in active learning (Dodson et al., 2018), peaks of activity in video learning (Kim et al., 2014), or how in-video prompting helps to prevent student disengagement in video learning (Shin et al., 2018). Numerous research focuses on collaborative aspects, such as designing online learning platforms (Alper et al., 2017) or tools that allow kids to create audiovisual projects on mobile devices (Hickey, 2019). Many tools are designed for collaborative work; however, not primarily in educational contexts. For example, tools are designed to collaboratively analyze rather specific audiovisual media artifacts, such as TV debates (Carneiro et al., 2019), or to engage in specific analytical activities, such as concept mapping in video learning (Liu et al., 2018), or collaborative video editing (Merz et al., 2018). The few collaborative educational tools mostly aim at peer learning through simple but asynchronous video annotation (Baecker et al., 2007; Singh et al., 2016) or at collaborative collection, creation, and assemblage of audiovisual media (Hamilton et al., 2018; Heimonen et al., 2013). A notable exception is the work by Chen, Freeman, and Balakrishnan (2019) who developed and evaluated a live streaming interaction tool for language teachers. Their study shows that adding video to language learning improves feedback and increases student engagement.

# 3 Methodology and Research Process

The development of TRAVIS GO was guided by a design-based research (DBR) approach (Design-Based Research Collective, 2003; Wang and Hannafin, 2005). This is a common method in learning sciences to study complex systems and to generate solutions to a problem that is then tested and evaluated in practice (Barab and Squire, 2004; Cobb et al., 2003). In our case, we identified the lack of adequate educational tools that also mind the didactic and methodological context as the problem, the design process of TRAVIS GO illustrates the implications how this problem was solved.

The development of TRAVIS GO primarily focused on teachers' needs and demands for an adequate analytical tool as they are the more critical population when it comes to adapting to and introducing new tools into teaching. Teachers are generally less open to use digital tools or web apps because audiovisual media artifacts are not a main subject in most school subjects but an additional perspective, for example, when comparing a book with a film in *Language* classes. Any new educational tool requires teachers to invest prior time and effort to familiarize with its features, to design didactic use cases that are adequate for a subject, grade, and curricula demands, and to make sure they understand a tool in a way they can teach and support students in working with it. Yet, there are little to no trainings offered to help teachers, and curricula provide no guidelines of how to adapt new tools into existing teaching practices. This leads to reluctance towards new tools and rather results in teachers using apps they know but are inapt. Teachers as well worry they could lose their authority as educator if they are not able to explain and demonstrate digital educational tools to students. In contrast, students are digital natives and more skilled in using and understanding digital tools, platforms, and digital media than teachers because of their high everyday use of, for example, social media and smart digital devices. Therefore, we define teachers as the main group to target when developing a web app for annotating and analyzing audiovisual media artifacts in educational context, especially because teachers rather than students are the gatekeepers who decide if a new tool will be used in class. It is above all important to make sure digital educational tools are actually used to teach basics of how to analyze and annotate audiovisual media artifacts. This means providing tools that appeal to teachers and fit into teaching routines.

We first researched *educational needs and demands* for video annotation tools by analyzing curricula for the above-mentioned subjects. In the development of TRAVIS GO, we used the school subjects of *Languages*, *Music*, *Arts*, and *History* as examples because these commonly involve teaching with and about audiovisual media. Second, we conducted *semi-structured expert interviews* with 31 teachers. Curricula define learning goals, interactions, and learning content for a school subject across grades. Teachers are experts who can provide first-hand insights into didactic practices and working with digital tools. Figure 1 describes the number of each type of data we collected by subject, state, and grade.

| | curricula | expert interviews | iterative testing feedback | video observation | student survey | teacher feedback |
|---|---|---|---|---|---|---|
| number | 30 | 31 | 11 | 5 | 105 | 5 |
| subjects | German | | | | | |
| | French | | | | | |
| | | English | | | | |
| | | Music | | | | |
| | | Arts | | | | |
| | History | | | | | |
| states | Basel-City | | | | | |
| | Basel-Country | | | | | |
| | | Bern | | | | |
| | Baden-Wuerttemberg | | | | | |
| | Lower Austria | | | | | |
| grades | 5-12 | | | | 10-12 | |

Figure 1. This figure chronologically (columns from left to right) shows the type and number of data we collected for each school subject, in each state, and for each grade (e.g. we reviewed 30 curricula in all subjects, for all states for grade 5-12.

We used qualitative document analysis (Flick, 2018) to extract all passages from the 30 curricula documents that define how audiovisual media artifacts should be integrated in *Languages, Music, Arts*, and *History* on different grades to identify key artifacts and key activities in each subject to be covered by the analytical features of the planned web application.

The 31 semi-structured expert interviews[1] (22 male, 9 female) (Bogner, Littig and Menz, 2009) provided rich information on processes and linked contexts (Meuser and Nagel, 2009), that is, needs and demand for collaborative video annotation tools. The interview questions covered motivations to use audiovisual media artifacts in teaching, application of educational standards to include audiovisual media artifacts into teaching, access to and use of technical equipment, current use of didactic audiovisual material, and expectations for a video annotation tool. In the qualitative thematic content analysis of the interview data (Flick, 2018), two experienced researchers generated initial codes through inductive coding for one interview with the software MAXQDA, compared codes, and agreed on a coding scheme. Then, one researcher proceeded to code the interview data by following the coding scheme. This process is an established way in the CSCW community to ensure validity of qualitative data through a high level of agreement (McDonald et al., 2019). The codes were then structured into nine categories: (1) *personal motivation to use audiovisual media in teaching*, (2) *institutional obligation to use audiovisual media in teaching*, (3) *technical and infrastructure aspects of teaching*, (4) *implemented didactic use of audiovisual media artifacts*, (5) *desired didactic use of audiovisual media artifacts*, (6) *implemented forms of work in teaching*, (7) *desired forms of work in teaching*, (8) *demands for using video annotation tools*, and (9) *critical assessment of using video annotation tools*. These categories served as basis for identifying key challenges in the development of TRAVIS GO (see 4) and informed the design of the web app.

---

[1] All interviews were recorded by permission, anonymized, and transcribed. All interviews were conducted in German, all quotes have been translated into analogous English.

According to DBR, in the process of designing and redesigning the first version of TRAVIS GO, we performed iterative tests with 11 teachers and collected their feedback through email and in conversations to evaluate the didactic and collaborative value of the web application. Because teachers only addressed minor aspects in mostly positive feedback we did not perform interventions but refined the first app version for the second and final version of TRAVIS GO.

# 4 Challenges to Address in the Development of TRAVIS GO

Analyzing curricula and expert interviews with teachers resulted in key challenges that were addressed in the development of TRAVIS GO in order for the web application to adequately meet user needs. We are foremost considering needs and demands of teachers as the crucial group of users. They are the gatekeepers who decide which tools they feel comfortable with to include into their teaching routines, therefore the premise is to develop a tool that improves accessibility and usability for teachers.

## 4.1 Curricula recommendations for integrating audiovisual media artifacts

The curricula analysis shows a general direction towards active and conscious ways to include various media forms and genres. Sampled curricula define similar standards for teaching analysis and interpretation of audiovisual media artifacts (see Figure 2), for example: analyzing films and commercials in relation to students' life world (*Arts*) (Erziehungsdepartement Basel-Stadt, 2013), using digital tools to describe film music (*Music*) (Ministerium für Kultus, Jugend und Sport Baden-Württemberg, 2016), interpreting film adaptations regarding cultural and historical contexts (*Languages*) (Erziehungsdirektion des Kantons Bern, 2017), or discuss the manipulability of language for political propaganda purposes in historical documentaries (*History*) (Erziehungsdepartement Basel-Stadt, 2013). Our interviews show teachers generally agree with curricula guidelines and requirements (*desired forms of work in teaching*) and are eager to follow curricula as institutional precepts (*institutional obligation to use audiovisual media in teaching*). However, no curriculum explicitly names digital tools for analyzing or annotating audiovisual media artifacts.

## 4.2 Motivation of teachers to engage with audiovisual media artifacts

German studies show teachers rarely use digital media and apps in teaching (18%), although they feel it would increase students' motivation to learn (88%) (Rohleder, 2019). Austrian studies find only 33% of media literacy classes actually discuss or interpret (audio-)visual media artifacts (Bundes Jugend Vertretung, 2017). In the expert interviews, teachers confirmed that students appreciate activities different from regular lessons and working with tools and platforms, like Moodle: *"For me, these are valuable tools, easy to learn and they are also motivating for students"* (English teacher, Basel-City). Teachers also said showing films or videos loosens the educational setting and improves teacher-student-interaction. However, we found teachers are primarily motivated by personal interest in, e.g. films, to further engage with audiovisual media artifacts in teaching (*personal motivation to use audiovisual media in teaching*) rather than by didactic examples, educational benefits, or suggested use cases provided by curricula (*desired didactic use of audiovisual media artifacts*).

| Subject | Artifact | Activity |
|---|---|---|
| *Arts* | Films<br>Commercials | reflect<br>describe<br>analyze<br>interpret |
| *Music* | Sound Phenomena<br>Film Music<br>Visual Art | identify<br>describe<br>analyze |
| *Languages* | Films<br>Film Adaptations<br>Music Videos | analyze<br>interpret<br>compare<br>understand |
| *History* | Historical Films<br>Documentaries | discuss<br>compare<br>interpret |

Figure 2. Overview of the media artifacts and the activities students should learn in each of the subjects according to curricula we analyzed.

## 4.3 Common ways of teaching with and about audiovisual media artifacts

The analysis of educational needs and demands for video annotation tools revealed that teachers in *Languages*, *Music*, *Arts*, and *History* mainly use audiovisual media artifacts to illustrate facts and contexts. Expert interviews showed teachers most commonly illustrate other learning resources with additional audiovisual media artifacts or guide students in producing videos in various contexts. However, both activities do not teach students, for example, to understand audiovisual coherences (*implemented didactic use of audiovisual media artifacts*). Analyzing audiovisual media artifacts is subject-specific, for example, comparing filmic presentation of characters to literary sources in Language classes: *"We look at the film and ask: What is the difference to the book? But also, what role does music play because you don't have that in the book?"* (English teacher, Basel-City). But teaching analysis is limited to teachers' knowledge of film analysis methodology and programs or apps they feel comfortable with, such as GarageBand, which often were designed for different purposes. Annotating films, videos etc. is the easiest for teachers and students to approach further analysis of audiovisual media artifacts, yet it is the least common. If used in teaching, associative annotation aims at teaching students to access their cultural memory through audiovisual media content: *"In my lessons I often work associatively and try to find out how much their* [the students'] *memory of associations is already filled"* (Art teacher, Basel-Country). The interviews revealed that work scenarios (*implemented forms of work in teaching*) are rather focused on teaching with than about audiovisual media artifacts.

## 4.4 Organizational and infrastructure aspects of teaching with audiovisual media artifacts

In interview teachers explained they prefer easy-to-use digital tools to avoid a disproportionate amount of time needed to adapt existing learning material versus time assigned to each subject per week: *"In the end I have very, very little time for my lessons and foremost I need to look at what's the outcome"* (History teacher, Basel-City). On the bright side, in contrast to German studies (Rohleder, 2019), we found that all our test schools in Switzerland had excellent technical infrastructure (fast wireless internet, computer rooms or mobile laptops, iPads, school server) to teach video analysis and annotation (*technical and infrastructure aspects of teaching*); however, teachers were missing adequate digital tools that meet their didactic needs.

## 4.5 Recommended educational tools for video analysis and annotation

A brief review reveals commonly-used tools and tools recommended by educational institutions (International Society for Technology in Education, 2017; Vega and Robb, 2019) are often not primarily designed for educational use and therefore limited in teaching students how to analyze and interpret media artifacts Recommended tools are often scientific and too complex (e.g., ELAN, VideoANT, Videonot.es), educational versions of media production or editing programs (e.g., Final Cut, Audacity, Loopmash), gamified programs (e.g., ArKaos, jam2jam, Songs2see), or designed for only specific school subjects and learning activities (e.g., Better Ears, Lilypond, Sibelius) (Klug and Schlote, 2018). In addition, our interviews show that teachers do not necessarily always see the use or benefit in officially recommended tools and apps (*critical assessment of using video annotation tools*) though they generally see a need to include such tools into teaching if they better match their needs and demands (*demands for using video annotation tools*).

Overall, these key challenges demonstrate that the bigger need is to develop educational tools that appeal to teachers by fitting into institutionalized routines and not by providing more complex features or approaches or new educational designs. In order to improve the accessibility and usability of educational tools, the development needs to address these educational challenges on organizational and institutional levels rather than on analytical or methodological levels.

# 5 Development and Design of TRAVIS GO

Based on the challenges identified in the analysis of curricula and interview data, we developed and designed the interface and features of TRAVIS GO to meet educational needs and demands as well as teaching routines (Schlote, Klug, and Neumann-Braun, 2020). The goal was to design a video annotation tool that is easy to understand, and allows to be quickly applied within established didactic routines and technical infrastructures. This is mostly based on the fact that teachers have only limited time per lesson and rely on well-practiced teaching routines. Therefore, accessing the web application should not lead to technical difficulties or take away valuable learning time. The web application should be free to use without time or location restrictions and be compatible with all digital devices, operating systems and browsers.

## 5.1 User Access

In the interviews, teachers expressed the need for a digital educational tool that reduces effort in preparing and teaching lessons with audiovisual media artifacts and does not cause additional technical or time effort: *"To be honest, set up and onboarding should take max five minutes"* (History teacher, Basel-City). Therefore, we designed TRAVIS GO as a web application that is accessible free of charge and without restrictions such as time or location. It is compatible with all digital devices, operating systems and browsers. Users only need to choose a temporal user name but do not need to register or login to use TRAVIS GO. In this way, the app reduces user and data management effort for the benefit of increasing openness, usability, inclusiveness, and fast onboarding to match organizational preconditions and didactic needs in school contexts. TRAVIS GO is available in German, English, French, and Italian. These and the following (5.1.1, 5.1.2) design implications help to solve the organizational challenges that teachers generally face in accessing and integrating digital educational tools into teaching routines.

5.1.1 Privacy and Data Storage

TRAVIS GO does not store any data and requires no registration or login but only a one-time user name. This absolves data security and privacy concerns that are exceptionally crucial for developing browser-based educational web applications in school contexts (Kumar et al., 2019), for example, when students and teachers use cloud services, such as Google Drive (Arpaci et al., 2015). Moreover, because TRAVIS GO users may be minors waiving data collection also avoids potential privacy issues in favor of inclusiveness.

| educational context | organizational challenge | design solution |
|---|---|---|
| lesson | little time | no user or data management |
| technical equipment | diverse & restrictive | no installing no desktop version |
| student data | sensitive (minors) | no login no account |
| audiovisual material | mind copyright | no data server |

Figure 3. TRAVIS GO provides design solutions to organizational challenges teachers face when using video annotation tools in educational context.

5.1.2 Copyright and Video

TRAVIS GO also does not store any media data but only relays media through the browser, evading any possible copyright issues concerning the use of films, YouTube videos etc. as online teaching media resources. Figure 3 shows how these organizational challenges for each educational context were solved in the design process of TRAVIS GO.

## 5.2 Project Management

In interviews, teachers unanimously expressed demand for an easy- and intuitive-to use tool with little to no learning curve. They voiced concern about losing authority if they are not able to quickly explain how to use the tool and its features and provide technical support to students. As a solution, to start a project in TRAVIS GO, users simply post the URL of the video or audio source (see Figure 4). To open a project, users drag-and-drop or choose the saved text file (see 5.5) from the hard drive. If users want to join a collaboration, they paste the collaboration code instead of a link (see 5.4).

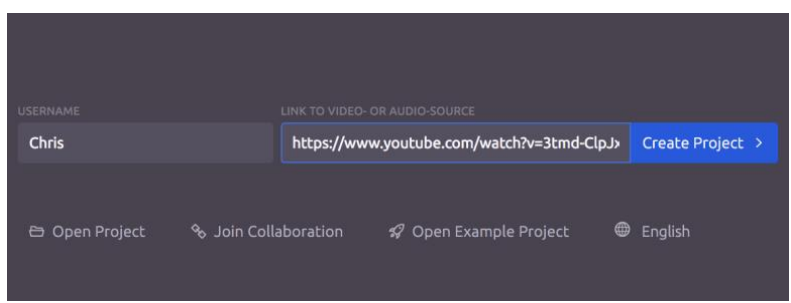

Figure 4. To start a new project in TRAVIS GO, users only need to paste a video URL (e.g. YouTube, Dropbox) on the app start page and give a temporary user name.

## 5.3 Work Space and Features

The interviews showed that teachers above all value accessibility, usability, and open collaborative work environments in order to appeal to students' media skills: *"In my classes we developed a very open exchange, a structure where as a teacher I'm almost dropping out. I find that very exciting"* (Arts teacher, Basel-Country). Within an openly structured work space, teachers demanded simplified analytical features compared to tools they currently used. Many teachers lack educational knowledge to teach film analysis *("The problem is missing education and knowledge in film science"*, German teacher, Basel-Country) or focus on filmic aspects, such as montage or camera movements, but not music and sound *("Music is only analyzed in a reduced way, there's no time or expertise for it"*, Art teacher, Bern). Figure 5 shows how the demand conducted in interviews played into the design of the TRAVIS GO work space and features. The *video player* (1) allows to play and navigate the selected video or audio source. In the *text editor* (2), users create annotation posts by setting a start and end point for a video sequence, and assigning an analytical category (picture, audio, text, meta) to the sequence. Each post shows up chronologically in the *annotation feed* (5) giving the drafted text and the timestamp, category, and username in a specific syntax ([00:12:34 PICTURE @username]). Projects can be given a *title and a description* (4), they can be *searched by text* (6), *shared as text file* (7), or in a *live collaboration* through a unique collaboration ID (8) (see 5.4). Projects can be *exported and saved* as Word or text document (9) (see 5.5). The *user list* (3) shows who is currently active (in green) in or contributed (in grey) to the project. Analytical features in TRAVIS GO are universal, they derive from film analysis (Bordwell, 2004) and allow to describe and analyze characteristics and effects of audiovisual media artifacts. The feature design enables subject-specific and case-sensitive adaptation without being technically deterrent for users. This methodological design allows for advancing common ways of teaching by providing a set of general analytical features that match teachers' desired didactic use of audiovisual media artifacts.

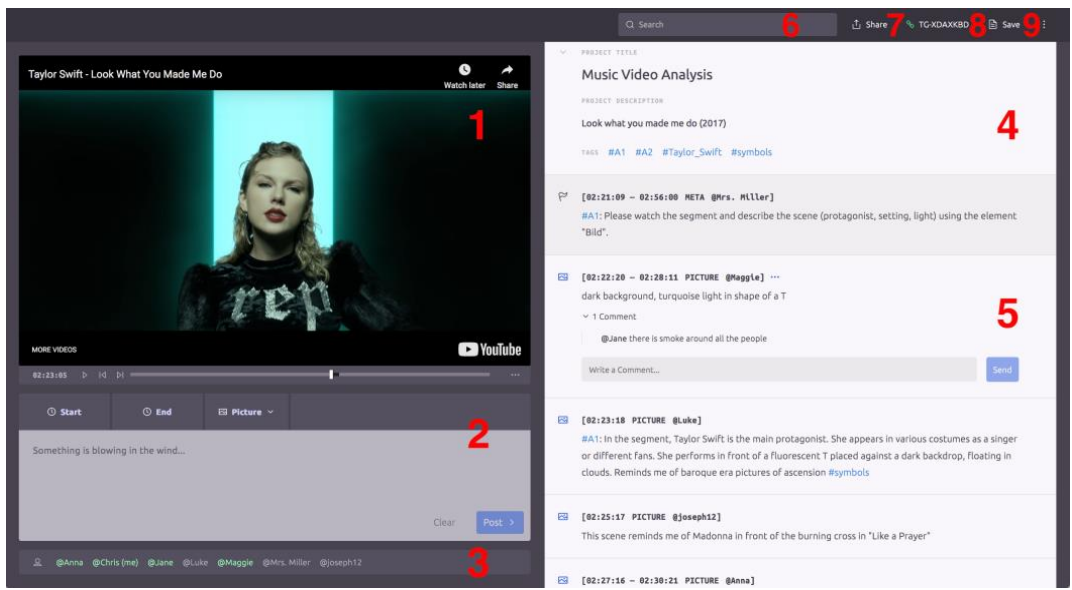

Figure 5. The TRAVIS GO work space provides a clear design and reduced features to match didactic requirements for teaching video annotation. The main work space areas are the video player (1), the text editor (2), and the annotation feed (4, 5).

Figure 6. This detailed view of the annotation feed in TRAVIS GO shows the assignment posted by the teacher (1) and posts by students (2), the hashtags students to structure their answers (3) and the list of hashtags (4).

5.3.1 Tagging, filtering, editing, and commenting on posts

Figure 6 shows a more detailed view of the annotation feed in TRAVIS GO. When writing a post in the text editor, users can use the #-symbol to tag words (e.g. #house). Tags appear as list in the project description and as auto-complete suggestions in the text editor. Timestamp, category, username, and hashtags can be clicked to filter posts, multiple filters can be applied. Each post can be edited by the post author. Each user can comment on any post. Tags enable associative exploration and subsequent detailed analysis of audiovisual media artifacts. Users can use predefined tags and add new tags to structure their results. Tags (#) and user handles (@) also allow for references in annotating and provide teachers with a simple and efficient overview on students work. Filtering in the annotation feed allows to display posts by results or to review results of single students since multiple students are able to collaborate on one project together (see 5.4). Filters help to evaluate progress and performance of individual students and learning results, for example, did students find and tag all scenes with a predefined tag. The basic interactive features in TRAVIS GO cater to curricula recommendations of how to teach about audiovisual media artifacts. They furthermore meet methodologic approaches teachers are already familiar with and improve the ways to work with audiovisual media artifacts.

5.3.2 Creating assignments and reviewing annotation results

To create assignments, teachers can use the meta category in the text editor to post one or more assignments in a project. They can define tags to indicate key words or analytical dimensions as part of assignments, for example, *"Describe the #camera_movement and #lighting in the opening scene"* or as presets for students to match with video sequences, for example, *"#closeup, #medium_shot, #long_shot"*. Filtering students' work can help teachers to consider results for grading classes. Teachers can assign students to comment on each other's posts to encourage peer-to-peer feedback, discussions or collaborative work which is a demand among teachers: *"Formal criteria are quite easy to teach, I guess, but how do I instruct students to reflect on it?"* (English teacher, Basel-City). Comments also enable teachers to give feedback with regard to students' individual contributions in group assignments.

## 5.4 Collaboration Mode

The analysis of teachers' educational needs and demands showed a strong demand for digital tools that allow better forms of collaborative work in peer-to-peer and student-to-teacher interaction. Teachers said, for example: *"We generally have difficulties in reaching the level of cooperative learning, going beyond working in groups. What I mean is collaborative knowledge development"* (Language teacher, Basel-City). TRAVIS GO supports collaborative work and exchange between students and teachers and students in a straightforward way (Klug, Schlote, and Eberhardt, 2017). The collaboration mode enables to initiate discussion in the web application and in face-to-face classroom interaction and allows teachers and students to give feedback on students' work. Multiple users can easily work simultaneously in the same project. Users generate a unique collaboration code (see Figure 7) that stays active as long as one user is active in the project. The collaboration code is displayed in the TRAVIS GO header (see Figure 8) and can be passed on by email, text etc., verbally, or by the teacher writing it on a board. Any person can join the project by entering the code into the app start page. This form of collaborating enhances existing teaching practices and adds didactic value to content-related depth, internal differentiation and cooperative learning. In this way, TRAVIS GO enables teacher-student inter-action which is required and desired in curricula and by teachers to increase their general motivation to use a digital tool in teaching.

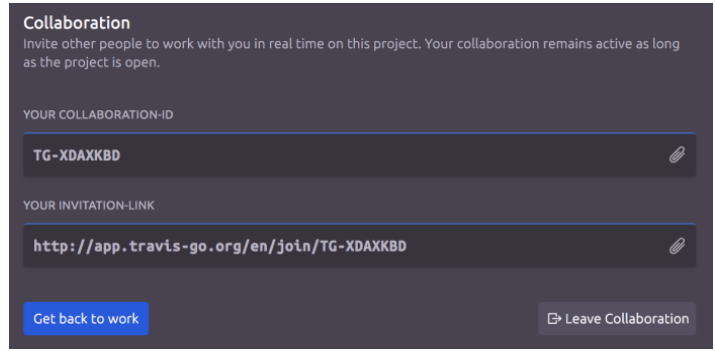

Figure 7. Any user can create a collaboration ID in a project at any time.

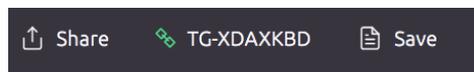

Figure 8. The collaboration ID can be shared via email, WhatsApp etc., verbally or by writing it on a board. The active collaboration ID is shown in the project menu bar.

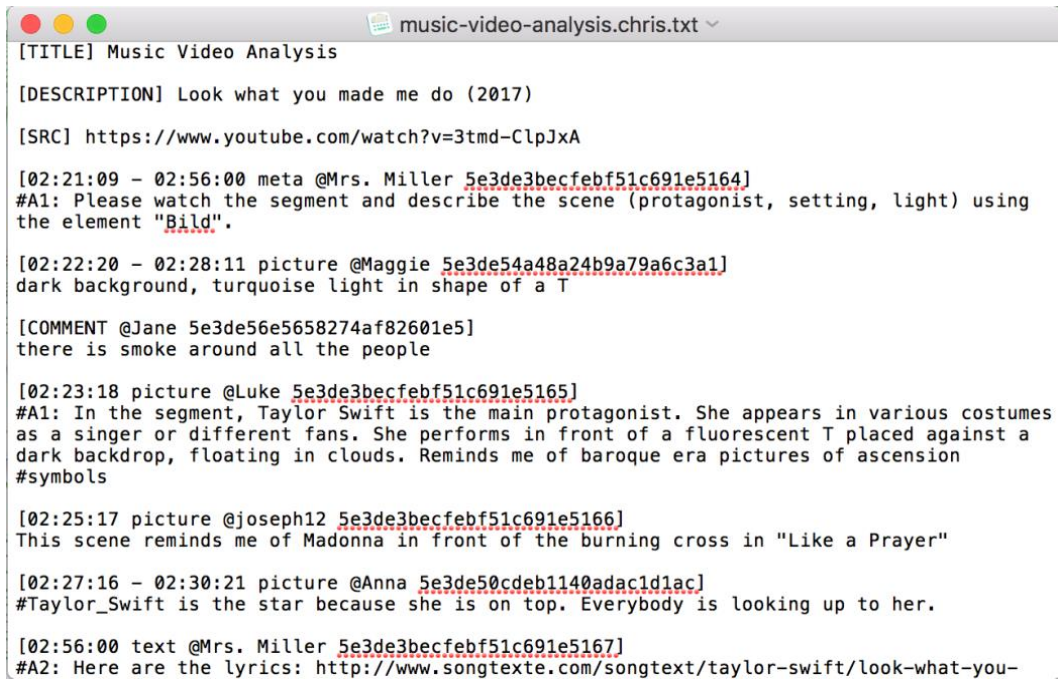

Figure 9. Any user in a project can save the project as text files.

## 5.5 Save and Share Projects

It is important, especially in educational contexts, that projects and analytical results can be easily exported from the web application into standard files. Because TRAVIS GO works without using a back-end data server, projects also need to be easily saved locally or in cloud services. In TRAVIS GO, users can save, load, and send projects as text files (see Figure 9). This meets schools' security standards using their own email and data servers. Users can also export projects as docx-files so students can create portfolios or hand them in as exams. Likewise, teachers can save projects as text files and share them with students or use them to start a collaboration for students to join (see 5.4).

# 6 Conclusion

We analyzed curricula for the school subjects *Languages*, *Music*, *Arts*, and *History*, and conducted 31 interviews with teachers in Switzerland and Germany. Curricula formulate the demand to teach analysis and interpretation of audiovisual media artifacts but do not give recommendations for adequate digital tools and provide only vague guidelines of how to integrate digital tools into existing teaching practices. From our interview study, we found the main problem to address are organizational limitations in teaching rather than the need for improved or more subject-specific analytical features or didactic environments. We found that an educational tool is much more beneficial for teachers when it is easy-to-use and fits into established teaching routines and that only basic analytical features are needed. The development of TRAVIS GO focused on designing a freely-available educational web application for simple and collaborate annotation of video and audio material that realizes these needs of teachers.

As a result, TRAVIS GO adds less value on methodological levels of educational video annotation tools in favor of openness and individual adaptability in subject-specific teaching contexts and by providing universal features for analyzing audiovisual media. The problem is that teachers are reluctant towards new digital tools if they are not comfortable using them, if they cannot adapt a tool into their teaching routines, and if a tool is too complex to easily use in little teaching time. TRAVIS GO solves these problems and adds greater value and innovation on organizational levels of integrating educational tools into established teaching routines. TRAVIS GO caters to teachers needs and concerns by reducing onboarding time, tool management, and time needed to familiarize oneself with its features and functionality.

- *TRAVIS GO provides an adequate analytical interface for educational use:*

The analysis of educational recommendations and educational needs and demands for video annotation tools revealed that teachers mainly use audiovisual media artifacts to illustrate facts and contexts (see 4.3) though they should teach analysis and interpretation (see 4.1). This is because of a lack of adequate tools and didactic approaches to perform more in-depth analytical discussions (see 4.5). TRAVIS GO supports this by providing an analytical interface that is action-oriented and allows exchange with others and to examine audiovisual media artifacts directly related to the material.

- *TRAVIS GO provides greater didactic freedom through reduced analytical structure:*

Teachers need an educational tool that provides subject-specific openness in creating assignments (see 4.3) and that can be integrated into established didactic settings without requiring additional effort from teachers (see 4.4). In TRAVIS GO, projects and assignments are not defined by the web application but can be designed according to subjects and learning goals to support a variety of didactic approaches. In order to offer added value beyond analog teaching-learning tools, digital learning tools should be customizable, interactive and adaptive. For TRAVIS GO, this has been realized by developing a reduced and open analytical structure that helps teachers to easily design learning activities around established didactic routines. TRAVIS GO is furthermore designed as a logical and consistent technical solution for didactic integration into existing teaching environments. This means, for example, it is up to teachers to choose audiovisual material, to assign tasks to groups or individuals or to decide to include students' contributions in TRAVIS GO into grading. Teachers as well need to supervise projects in TRAVIS GO and interact with students to coach them with their work.

TRAVIS GO was developed within educational context but is not limited to this purpose. We pointed out that our focus is to provide an easy-to-use web app that allows for a quick and straightforward integration into inevitable limitations of existing educational situations and not on designing new subject-specific analytical features or environments. Accessibility, usability, and adaptability to existing practices are more crucial keys for deciding to use a digital tool. Analytical dimensions and features are universal for analyzing and annotating any audiovisual format in any context and differ in their interpretative perspectives on each subject and context. TRAVIS GO provides a free workspace for individual specification based on universal analytical dimensions. This allows to use TRAVIS GO in any other context that involves forms of collaboratively reviewing, discussing, or annotating audiovisual media artifacts, such as video production, video evaluation, or higher education.

Overall, our results contribute further insights for the CSCW and the CSCL community about the need to include user demands into developing educational tools and how to implement these in the development and design process of a web application for collaborative video annotation. Schools and teachers are main agents in supporting students' media literacy. Teenagers frequently consume and interact with audiovisual media artifacts. But teachers lack adequate didactic tools and methodological skills to analyze audiovisual media artifacts according to curricula standards. TRAVIS GO demonstrates a successful way how to solve these issues and challenges in the design and development of an open educational resource (OER).

# 7 Limitations and Future Work

Our study is limited to that we did not evaluate students' needs and demands as part of designing TRAVIS GO. This is because we identified teachers as more crucial regarding openness, prejudices, and knowledgeability towards introducing and using new tools into existing teaching routines, therefore is it important to primarily evaluate teachers' attitudes, mindsets, and ideologies when designing educational tools. Our results are furthermore limited because we are not including data about the validation of TRAVIS GO in realistic educational environments. Although we tested and evaluated TRAVIS GO through video observations of students and teachers working with the app, data collection from using TRAVIS GO in various school subjects and didactic settings is still ongoing and part of future work.